\documentclass{article}
\usepackage{authblk} 
\usepackage{amsmath}
\usepackage{bm}
\usepackage{amssymb}
\usepackage[colorlinks=true, citecolor=black, linkcolor=black, urlcolor=black]{hyperref} 
\usepackage{float}
\usepackage[numbers,sort&compress]{natbib} 
\usepackage{caption} 
\usepackage{graphicx}

\usepackage{geometry}

\begin{document}
	
	\title{Quantum Generative Adversarial Networks in a Silicon Photonic Chip with Maximum Expressibility}
	\author{Haoran Ma}
	\author{Liao Ye}
	\author{Fanjie Ruan}
	\author{Zichao Zhao}
	\author{Maohui Li}
	\author{Yuehai Wang}
	\author{Jianyi Yang\thanks{yangjy@zju.edu.cn}}
	\affil{College of Information Science and Electronic Engineering, Zhejiang University, Hangzhou 310027, China}
	\date{}
	\maketitle
	
	\begin{abstract}
		Generative adversarial networks (GANs) have achieved remarkable success with realistic tasks such as creating realistic images, texts, and audio. Combining GANs and quantum computing, quantum GANs are thought to have an exponential advantage over their classical counterparts due to the stronger expressibility of quantum circuits. In this research, a two-qubit silicon quantum photonic chip is created, capable of executing arbitrary controlled-unitary ($C\hat{U}$) operations and generating any 2-qubit pure state, thus making it an excellent platform for quantum GANs. To capture complex data patterns, a hybrid generator is proposed to inject nonlinearity into quantum GANs. As a demonstration, three generative tasks, covering both pure quantum versions of GANs (PQ-GAN) and hybrid quantum-classical GANs (HQC-GANs), are successfully carried out on the chip, including high-fidelity single-qubit state learning, classical distributions loading, and compressed image production. The experiment results prove that silicon quantum photonic chips have great potential in generative learning applications.
	\end{abstract}
	
		\section{Introduction}
	Quantum computing is one of the most fascinating computing paradigms today because of its potential for exponential speedup compared to classical computers in certain tasks. For example, Shor's algorithm can factor enormous numbers in polynomial time\textsuperscript{\cite{1}}, whereas the HHL algorithm can considerably accelerate the solution of linear systems\textsuperscript{\cite{2}}. However, these algorithms require a quantum computer that is error resilient. Despite recent breakthroughs\textsuperscript{\cite{3}}, quantum devices are not yet capable of running these algorithms for practical applications. To utilize restricted quantum resources, the hybrid quantum-classical (HQC) framework has been proposed\textsuperscript{\cite{4,5,6}}. The main idea of HQC is to split tasks into two parts: one executed efficiently on quantum devices and the other completed on classical computers\textsuperscript{\cite{6}}. Therefore, the computational burden on quantum computers is reduced. Quantum machine learning (QML) is one of the beneficiaries of HQC, in which quantum devices are outfitted with parameterized quantum gates with trainable parameters, known as the parametrized quantum circuit (PQC), and a classical computer uses the measured results to adapt these parameters for learning. Both theoretical and experimental research indicates that QML may outperform its traditional counterpart\textsuperscript{\cite{7,8,9,10}}.
	
	Quantum GANs, a breakthrough method in QML, have gained extensive interest for their potential to exhibit quantum advantage on near-term quantum devices\textsuperscript{\cite{11}}. This method is made up of two models: the generator (G) and the discriminator (D), who are engaged in a minimax game in which D learns to discriminate between real and generated data while G strives to produce data that tricks D. Finally, G learns how to capture the distribution of the training dataset and create new data. Quantum GANs can be employed for generating both quantum states\textsuperscript{\cite{12,13,14,15,16,17}} and classical data\textsuperscript{\cite{18,19,20,21,22,23,24,25,26,27}}. In the case of generating quantum states, both G and D must be quantum, resulting in PQ-GAN. When generating classical data, either G or D is quantum in most cases, leading to HQC-GAN. In this paper, we show both forms of quantum GANs on a programmable silicon quantum photonic chip.
	
	Silicon quantum photonic is a promising quantum computing platform that uses single photons as quantum information carriers, employs optical waveguides to guide and route photons, and utilizes Mach-Zehnder interferometers (MZIs) to manipulate the photonic qubit, offering phase-stable quantum circuits, high-precision quantum operations, and scalability to large-scale implementations\textsuperscript{\cite{28,29}}. Recently, silicon quantum photonic chips have been employed to demonstrate quantum GANs. Ref. \cite{17} demonstrates the fault-tolerance of PQ-GANs in photonics by producing high-quality quantum states despite chip noise and imperfections. However, their generator is only capable of creating two-photon maximum entangled states, limiting the expressive power required for generative models. Ref. \cite{27} adopts HQC-GAN as a submodule for financial applications, where the  quantum GAN aims to load the financial data into the quantum state of only one photon. Yet, loading data into multi-photon quantum states may be more meaningful, as entanglement is important for realizing quantum advantage\textsuperscript{\cite{30}}. Therefore, designing a chip capable of generating arbitrary multi-photon quantum state is crucial for quantum GANs.
	
	In this work, we suggest a scheme to produce on-chip arbitrary two-qubit pure states by combining asymetrical MZIs (AMZIs), frequency post-selection, and $C\hat{U}$ operation. As a demonstration, a silicon photonic chip is fabricated and used to perform three generative tasks, including both PQ-GAN and HQC-GAN. Compared with Ref. \cite{17}, our quantum photonic generator is maximally expressive. In addition, a hybrid generator consisting of classical and quantum layers is proposed to introduce nonlinearity into quantum GANs, thereby capturing more complex data features than the pure quantum version. We start by demonstrating a PQ-GAN, training the device to produce single-qubit pure and mixed states with fidelities as high as 99.41\% and 98.39\%, respectively. Then, we employ an HQC-GAN to undertake approximately distribution loading, where the data is loaded into the basis occurance probabilities of a two-photon quantum state rather than single-photon in Ref. \cite{27}. Finally, we show how to generate compressed MNIST images with a modified HQC-GAN, where the proposed hybrid generator is utilized. To the best of our knowledge, this is the first proof-of-principle demonstration of quantum GANs to learn mixed quantum states and produce images using a silicon photonic chip.
	
	\section{Silicon Photonic Chip for Quantum GANs}
	\subsection{The AMZI with Frequency Post-selection}
	AMZI is an indispensable component of our chip, whose combination leads to the generation of amplitude-adjustable entangled states, which is crucial for our scheme. In this subsection, we describe the transformation that AMZI applies to the quantum state of photon pairs under post-filtering, which results in the measured coincidence rate (CR) varying with the phase changes of AMZI. The relationship between CR and AMZI's phase angle $\varphi$ is theoretically derived and experimentally verified. 
	
	We begin with theoretical derivations. \textbf{Figure 1a} displays a typical AMZI, where signal/idler photon pairs at different wavelengths are injected from port $a$ and split to port $c$ and $d$. The transmission matrix of the AMZI is
	
	\begin{equation}
		\begin{aligned}
			U_{AMZI}&=
			\begin{bmatrix}
				1 & i \\
				i & 1
			\end{bmatrix}\cdot
			\begin{bmatrix}
				e^{i(\beta_{s/i}\triangle l+\varphi)} & 0 \\
				0 & 1
			\end{bmatrix}\cdot
			\begin{bmatrix}
				1 & i \\
				i & 1
			\end{bmatrix}\\
			&=e^{i\frac{\beta_{s/i}\triangle l+\varphi+\pi}{2}}
			\begin{bmatrix}
				sin((\beta_{s/i}\triangle l+\varphi)/2) & cos((\beta_{s/i}\triangle l+\varphi)/2) \\
				cos((\beta_{s/i}\triangle l+\varphi)/2) & -sin((\beta_{s/i}\triangle l+\varphi)/2)
			\end{bmatrix}
		\end{aligned}
	\end{equation}
	where $\beta_{s/i}$ is the transimission constant of the signal/idler photon and $\triangle l$ is the length difference of two arms. Assuming the free spectral range (FSR) of AMZI is designed to be twice the wavelength difference of photon pairs, we have $|\beta_s\triangle l-\beta_i\triangle l|=\pi$. Therefore, the transmission matrices of the signal and idler photons can be expressed as
	
	\begin{equation}
		U_{s}=e^{i\frac{\varphi+\pi}{2}}
		\begin{bmatrix}
			sin(\varphi/2) & cos(\varphi/2) \\
			cos(\varphi/2) & -sin(\varphi/2)
		\end{bmatrix},
		U_{i}=e^{i\frac{\varphi+2\pi}{2}}
		\begin{bmatrix}
			cos(\varphi/2) & sin(\varphi/2) \\
			sin(\varphi/2) & -cos(\varphi/2)
		\end{bmatrix}
	\end{equation}
	
	The initial state of photon pairs can be written in terms of the creation operator on the vacuum state as $a^\dagger_sa^\dagger_i|vac\rangle$. The AMZI performs $U_s$ and $U_i$ for signal and idler photons, respectively, where the process can be described as
	
	\begin{equation}
		\begin{aligned}
			a^\dagger_sa^\dagger_i|vac\rangle & \stackrel{U_{AMZI}}{\longrightarrow}
			e^{i\varphi+\frac{\pi}{2}}(sin(\varphi/2)c^\dagger_s+cos(\varphi/2)d^\dagger_s)\cdot(cos(\varphi/2)c^\dagger_i+sin(\varphi/2)d^\dagger_i)|vac\rangle \\
			&=e^{i(\varphi+\frac{\pi}{2})}\{sin(\varphi/2)^2c^\dagger_sd^\dagger_i+other.\}|vac\rangle
		\end{aligned}
	\end{equation}
	
	By post-selecting the case $c^\dagger_sd^\dagger_i$ with filters, the success probability of coincidence is $sin^4(\varphi/2)$. If the photons enter from port $b$, the result becomes $cos^4(\varphi/2)=sin^4((\varphi+\pi)/2)$. Thus, we always have
	
	\begin{equation}
		C(\varphi)\propto C_{max}sin^4(\varphi/2)	
	\end{equation}
	where $C_{max}$ is the measured maximum CR, which occurs when signal and idler photons are deterministically separated. This result is also experimentally verified and shown in \textbf{Figure 1b}. The dots are measured CRs by tuning the two AMZIs' phases $\varphi_1$ and $\varphi_2$ in \textbf{Figure 2a}, respectively, while the solid lines are fitted results using $a sin^4(\alpha I^2+\beta)+b$.
	
	\begin{figure}[H]
		\centering{\includegraphics[width=0.6\textwidth]{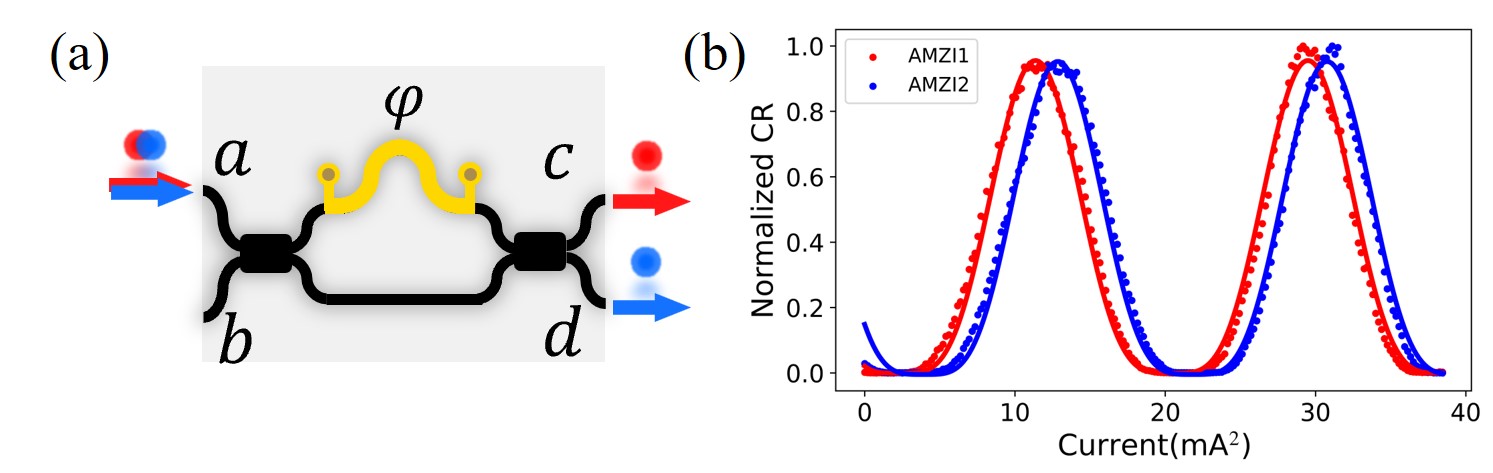}}
		\caption*{\textbf{Figure 1.} a)The scheme of AMZI. (b)The measured CRs vary with the square of the current applied to the two AMZIs.}
		\label{fig:boat1}
	\end{figure}
	
	\subsection{Chip Design and Experimental Setup}
	The quantum GANs are exhibited on a digital silicon quantum photonic chip, as shown schematically in \textbf{Figure 2a}. The chip is pumped with continuous-wave (CW) light at 1550.92 nm, which is amplified to 30 mW by an erbium-doped fiber amplifier (EDFA). After being adjusted by a polarization controller (PC) and filtered by a dense wavelength division multiplexer (DWDM), light is coupled into the chip via a grating coupler with a coupling loss of about 4.5 dB. Photon pairs are created in two 1-cm-long waveguide spiral sources using spontaneous four-wave mixing (SFWM) and routed through single-mode waveguides with 450nm$\times$220nm. Interferemeters, which include multimode interferometers (MMIs), beam-splitters and thermo-optic phase shifters (PSs) driven by a current source, are used to manipulate photonic qubits. The output photons are filtered using DWDMs, which selectively isolate signal photons at 1555.75nm (red) for the control qubit and idler photons at 1546.12nm (blue) for the target qubit. After polarization adjustments, photons are detected by superconducting nanowire single-photon detectors (SNSPDs) (Photec, 50Hz dark counts, 90\% efficiency). A TimeTagger (Swabian) is utilized to perform coincidence counting, with a coincidence window of 300 ps. The measured coincidence rate (CR) is approximately 3 kHz. A classical computer is used to automate the quantum GANs, encompassing quantum gate manipulation, data acquisition, instant analysis, and the implementation of classical neural networks (NNs).
	
	When pumped, the two spiral sources have equal power, resulting in the production of two-photon NOON state $(|2_{s,i},0\rangle +|0,2_{s,i}\rangle)/\sqrt{2}$. The photon pairs from two sources are filtered and partially split by two AMZIs, where the FSR of each AMZI is 19.20nm, equaling twice the wavelength difference of the photon pairs. The photons are then switched by a waveguide crossing, yielding the path-entangled state
	
	\begin{equation}
		|\psi\rangle_0 =e^{i\theta_8}\sqrt{C_1(\varphi_1)}|0_s\rangle|0_i\rangle+\sqrt{C_2(\varphi_2)}|1_s\rangle|1_i\rangle)/A
	\end{equation}
	where $A$ is the normalization coefficient, and $C_1(\varphi_1)$ and $C_2(\varphi_2)$ represent the measured CR of source1 and source2, respectively, which can be calculated using Equation (4). Due to the global phase introduced by AMZI in Equation (1), a relative phase occurs in the entangled state, which can be compensated for by adjusting $\theta_8$. The state can be further described as
	
	\begin{figure}[H]
		\includegraphics[width=1\textwidth]{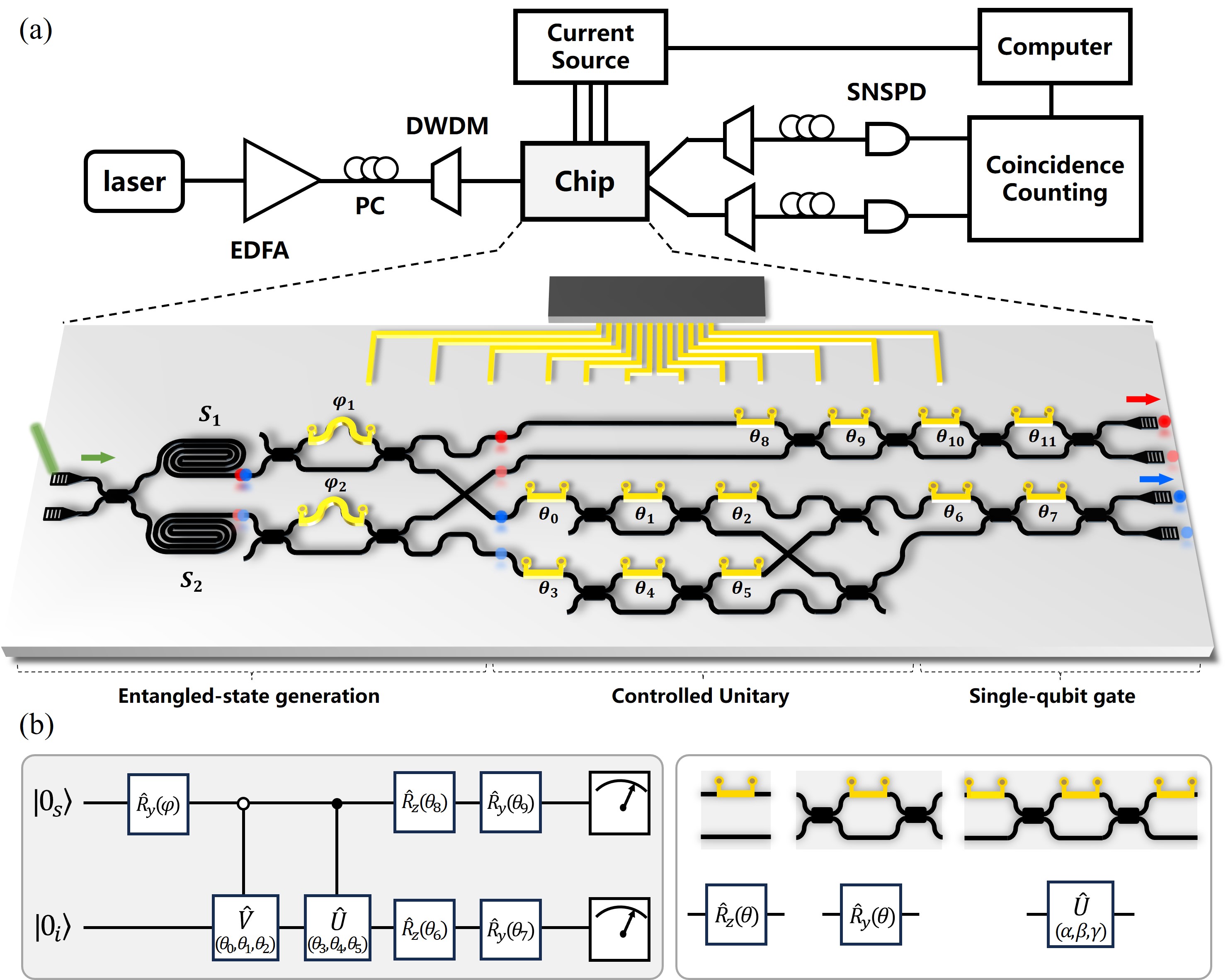}
		\caption*{\textbf{Figure 2.} a)Experimental setup and the programmable silicon quantum photonic chip. The chip is made up of three parts: i) creating two-photon entangled states with controllable amplitudes using SFWM and AMZIs, ii) performing controlled unitary operations to produce arbitrary two-qubit pure states, and iii) using single-qubit gates for computing or state tomography. b)The equivalent quantum circuit of the chip. Two waveguides with one equipped with phase shifter corresponds to an $\hat{R_z}$ gate, while the MZI corresponds to an $\hat{R_y}$ gate. The combination of both can carry out any SU(2) transformation.}
		\label{fig:boat1}
	\end{figure}

	\begin{equation}
		\begin{aligned}
			|\psi\rangle_0 &= e^{i\theta_8}cos(\varphi)|0_s\rangle|0_i\rangle+\sin(\varphi)|1_s\rangle|1_i\rangle\\
			&=\alpha|0_s\rangle|0_i\rangle+\beta|1_s\rangle|1_i\rangle
		\end{aligned}
	\end{equation}
	where the rotation angle $\varphi$ is determined by both the two tunable split ratios of AMZIs, with the relationship $tan(\varphi)=\sqrt{C_2(\varphi_2)/C_1(\varphi_1)}$. As a result, entangled state with arbitrary amplitude is generated. 
	
	Next, a $C\hat{U}$ operation is carried out using the approach in Ref. \cite{31}. The space of idler photon is expanded by adding two waveguide modes. The path of the photon introduces the third dimention indicated by $|0_p\rangle$ and $|1_p\rangle$, choosing which operation the idler qubit experiences — $\hat{V}$ or $\hat{U}$. The path is controlled by the state of the signal photon, resulting in a superposition of $\hat{V}$ and $\hat{U}$ operations
	
	\begin{equation}
		\alpha|0_s\rangle\otimes\hat{V}|0_i\rangle\otimes|0_p\rangle+\beta|1_s\rangle\otimes\hat{U}|0_i\rangle\otimes|1_p\rangle
	\end{equation}

	A waveguide crossing and two MMIs are then used to erase the path information. By coincidence, we have the state after $C\hat{U}$
	
	\begin{equation}
		|\psi\rangle_1=\alpha|0_s\rangle\otimes\hat{V}|0_i\rangle+\beta|1_s\rangle_T\otimes\hat{U}|0_i\rangle
	\end{equation}
	
	it can be rewritten in the computational basis, which is
	
	\begin{equation}
		|\psi\rangle_2=\sqrt{p_0}|00\rangle+\sqrt{p_1}|01\rangle+\sqrt{p_2}|10\rangle+\sqrt{p_3}|11\rangle
	\end{equation}
	where $p_0+p_1+p_2+p_3=1$ and the relative phases have been ignored. Obviously, the coefficient before each basis is arbitrary, thus the state can be any two-qubit pure state. This is a deviation from prior works\textsuperscript{\cite{32,33,34,35,36}}. Finally, additional PSs and MZIs are added as single-qubit gates to perform state tomography or other sophisticated operations.
	
	The equivalent quantum circuit is depicted in \textbf{Figure 2b}, correlating one-to-one with the chip's PS values, except for $\theta_{10}$ and $\theta_{11}$ because of broken PS $\theta_{11}$. However, this does not affect our experiments because the issue can be addressed by altering $\theta_{8}$ and $\theta_{9}$ (See Appendix A). The parameterized gate $\hat{R}_y(\varphi)$ is constructed by the two AMZIs and PS $\theta_8$ combined. The remaining $\hat{R}_z$ and $\hat{R}_y$ gates are built respectively using PSs added to the waveguides and the MZIs. In the quantum circuit context, the ability of our technique to generate arbitrary states is further confirmed by Ref. \cite{37}, which states that the combination of $\hat{R}_y$ gate on the control qubit and two unitaries controlled by $|0\rangle_C$ and $|1\rangle_C$ can prepare any two-qubit state. For the sake of simplicity, we will illustrate our chip's configuration in subsequent experiments using quantum circuit language.
	
	\section{Experiments and Results}
	\subsection{Learn Single-qubit State}
	Generating quantum states is vital in quantum computing\textsuperscript{\cite{30}}. We first exhibited the PQ-GAN, whose architecture is illustrated in \textbf{Figure 3a}. The black box provides quantum real data defined by a density matrix $\sigma$, yet the internal physical structure and quantum process are not exposed. The generator G, which corresponds to the quantum circuit inside the red dashed box in \textbf{Figure 3b}, can generate any quantum state (single-qubit in our case), including pure and mixed states. The mixed state is created by partially tracing the entangled state. D performs quantum measurements ($\mathcal{M}$) on both the real and produced states, intending to differentiate between them based on the statistical features of the measurement results $p_{\rho}=tr(\mathcal{M}\rho)$ and $p_{\sigma}=tr(\mathcal{M}\sigma)$. \textbf{Figure 3b} depicts D's quantum circuit as the portion within the blue dashed box. The final two gates of the signal qubit are unused, with settings set to perform the identity operation $I$.  The optimization mission of PQ-GAN is
	
	\begin{equation}
		\mathop{min}\limits_{\bm{\theta_g}} \mathop{max}\limits_{\bm{\theta_d}} tr[\mathcal{M}(\bm{\theta_d})\rho(\bm{\theta_g})]-tr[\mathcal{M}(\bm{\theta_d})\sigma]
	\end{equation}
	where $\bm{\theta_g}$ and $\bm{\theta_d}$ represent the trainable phases, which are initialized with normal distribution $\mathcal{N}(\mu=0,\sigma=0.2)$. During the training process, G and D are alternatively optimized until they converge. To eliminate the influence of random factors, for pure and mixed state learning, we trained with different and identical beginning parameters, respectively, for 5 rounds, with 200 epochs per round.  In each training epoch, D was trained for three steps first, then G for one. Gradient descent (ascent) is used to update the parameters. The detailed experimental parameters can be found in Appendix B.
	
	A pure single-qubit state is learned first. The real state is selected as $\sigma_p=(|0\rangle+|1\rangle)(\langle 0|+\langle 1|)/2$, which is prepared by setting $\hat{R}_y(\phi)=I$ and $\hat{V}=H$. The density matrix, depicted in \textbf{Figure 3d}, is acquired by quantum state tomography (QST). \textbf{Figure 3c} plots the curve of negative D loss, with the solid line representing the mean value and the shaded area representing the standard deviation (STD) over 5 rounds. The negative D loss keeps decreasing and approaches zero, implying that D is unable to discern the source of the state and that G has successfully replicated the genuine state. To track the resemblance between the produced and real states, we retrieved $\rho_p$ by QST and estimated the fidelity with $F(\rho,\sigma)=[Tr(\sqrt{{\sqrt{\rho}\sigma\sqrt{\rho}}})]^2$ at each epoch. As illustrated in \textbf{Figure3c}, $F$ approaches 1 after approximately 100 epochs. The final generated state is depicted in \textbf{Figure 3d}, with a fidelity as high as $99.41\pm0.54$\%. Such a high fidelity implies that our chip can easily learn pure states.
	
	\begin{figure}[H]
		\includegraphics[width=1\textwidth]{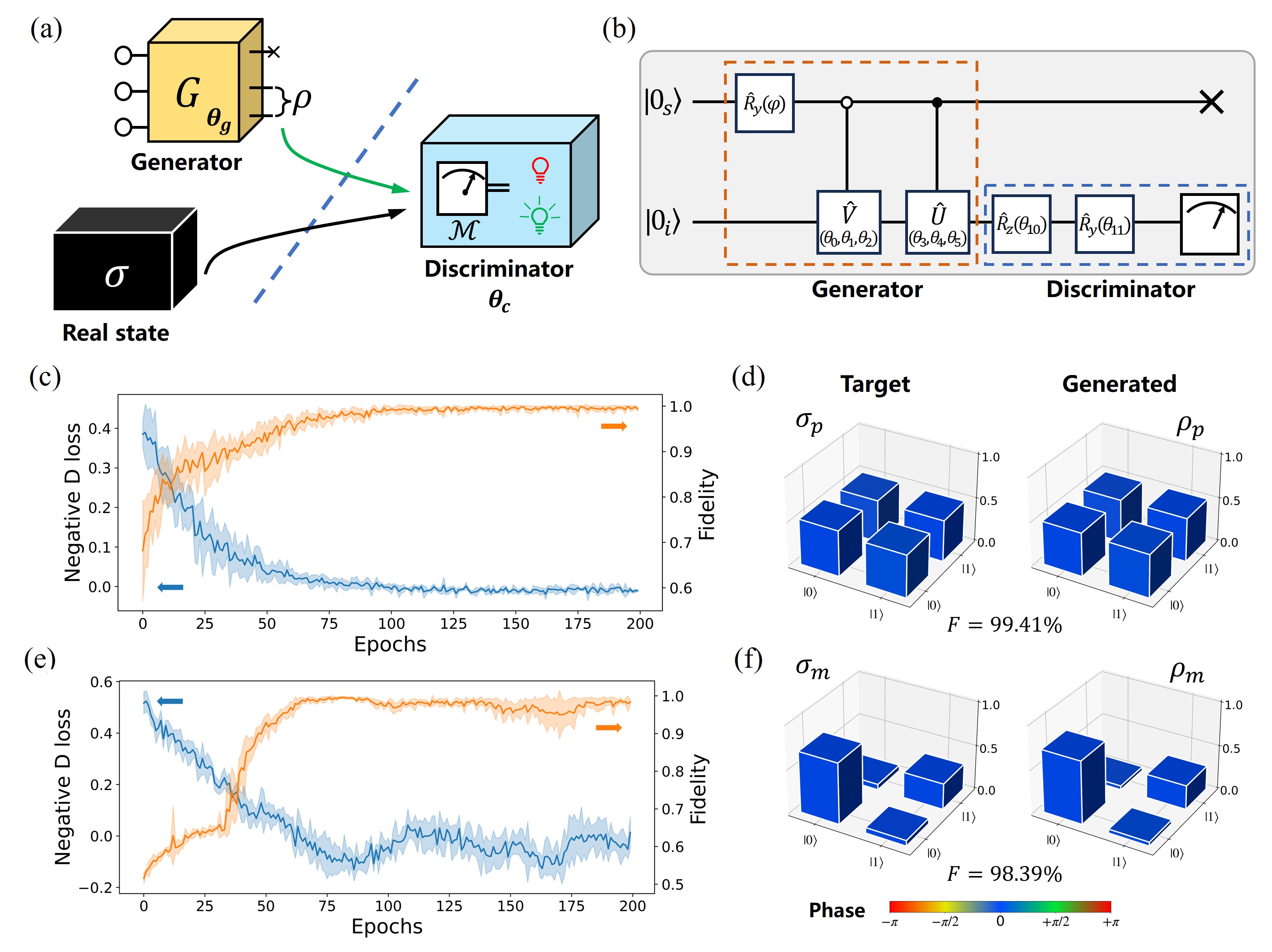}
		\caption*{\textbf{Figure 3.} a)The structure of PQ-GAN, where real data is a quantum state $\sigma$, G generates arbitral states $\rho$, and D does projective measurements $\mathcal{M}$. b)The components of G and D used in the chip, which are shown in the form of quantum circuits. c)The tracking of negative D loss and the fidelity of pure state learning over 200 epochs. d) The measured dense matrix of $\sigma_p$ and $\rho_p$ with a mean state fidelity of $99.41\%$. e)The mixed state learning process. f)The measured dense matrix of $\sigma_m$ and $\rho_m$ with a mean state fidelity of $98.39\%$.}
		\label{fig:boat1}
	\end{figure}
	
	Then, the learning of a mixed state $\sigma_m=(7|0\rangle\langle0|+3|1\rangle\langle1|)/10$ is investigated. The real state $\sigma_m$ is created by randomly preparing $\{|0\rangle,|1\rangle\}$ with probabilities $\{0.7,0.3\}$. \textbf{Figure 3e} plots the evolution of negative D loss and fidelities, which oscillate around 0 and 1 after 75 epochs, respectively. This disconvergence of mixed state learning is known as limit cycles, which result from the bilinear nature of the loss function when using gradient descent to generate mixed states in adversarial training\textsuperscript{\cite{14}}. The occurance of bilinear problem may be due to the mixed state being positioned inside the Bloch sphere, where the nonlinear restrictions from the Bloch boundary are not satisfied\textsuperscript{\cite{14}}. We displayed $\sigma_m$ and the final generated $\rho_m$ in \textbf{Figure 3f}, with the fidelity $98.39\pm0.60$\%. The ability to demonstrate mixed-state learning comes from the fact that the chip can produce single-qubit states in any mixedness by tracing out two-qubit pure states. 
	
	\subsection{Load Classical Distribution}
	Quantum GANs can encode classical data into quantum states using a low-depth quantum circuit by training the generator, overcoming the data input bottleneck in QML\textsuperscript{\cite{19,20}}. Since the distribution is arbitrary, a generator with high expressivity is required. We demonstrate that our chip has the capabilities by performing three distribution loading tasks. To achieve this goal, an HQC-GAN architecture is adopted, and its diagram is depicted in \textbf{Figure 4a}. It is the quantum version of Wassertein GAN with gradient penalty (WGAN-GP)\textsuperscript{\cite{38,39}}, which employs Wassertein distance (or Earth mover's distance) as the loss function, alleviating
	
	\begin{figure}[H]
		\includegraphics[width=1\textwidth]{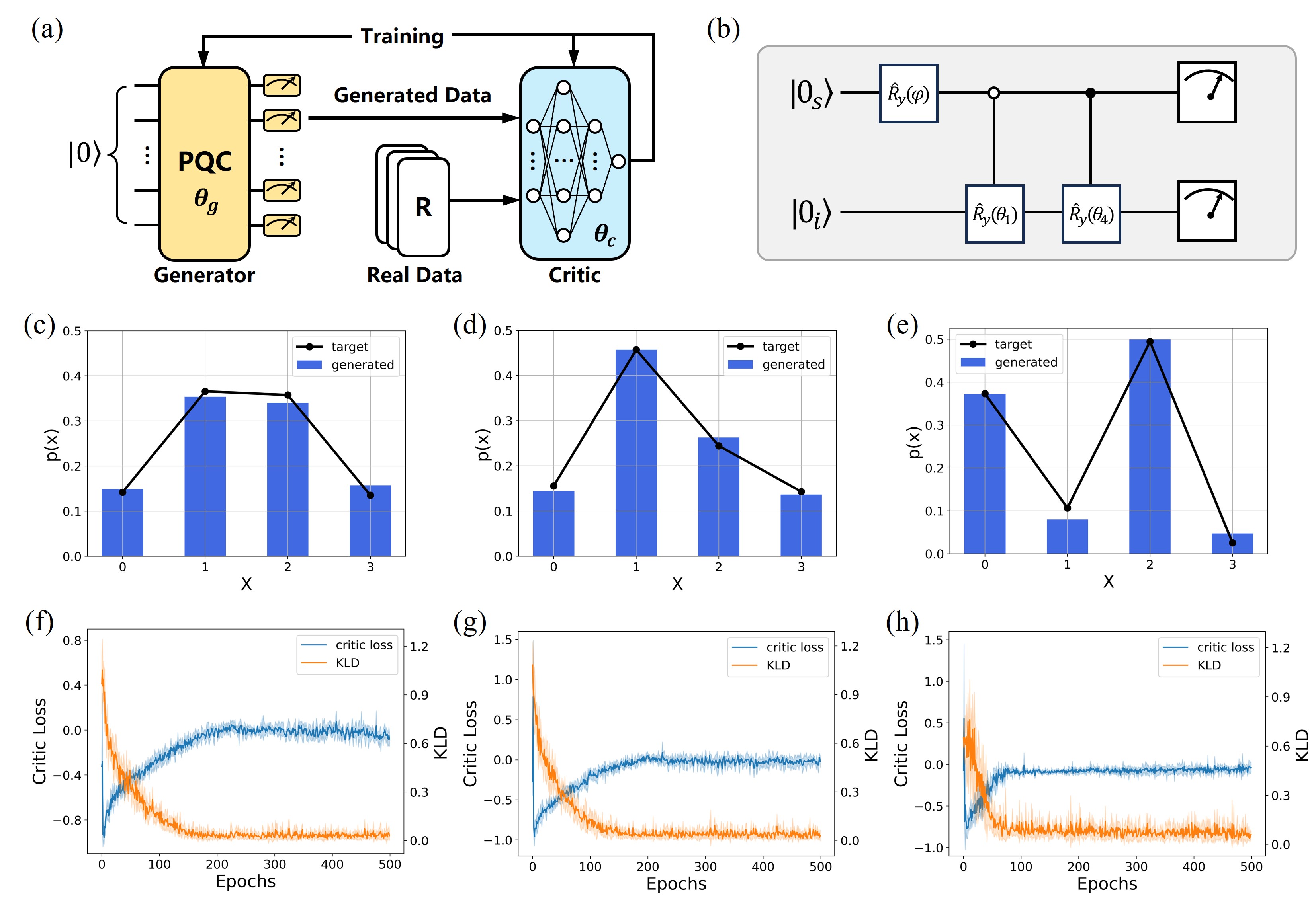}
		\caption*{\textbf{Figure 4.} a) The architecture of HQC-GAN, the quantum variant of WGAN-GP, where the generator is replaced by a PQC $U(\bm{\theta_g})$. b)G's quantum circuit as configured on the chip, with just three trainable parameters and the rest fixed to perform operation $I$. c)-e)Plots of the three distributions, which are the normal, log-normal, and bimodal distributions from left to right. f)-h) Curves of KLD (orange) and critic loss (blue) of three distributions across 500 epochs.}
		\label{fig:boat1}
	\end{figure}
	
	\noindent the issue of unstable training, mode collapse, and vanishing gradients in standard GAN\textsuperscript{\cite{40, 41}}.
	
	The HQC-GAN has two components: a quantum G and a classical critic. G is a PQC parameterized with $\bm{\theta_g}$, and the generated data is encoded into the occurance probabilities of basis states, which can be acquired by measuring along the computational basis. The measurement operation also incorporates the stochastic aspect required by G. In our experiment, the chip plays the role of G. In \textbf{Figure 4b}, the chip is set to realize a quantum circuit with only three trainable parameters, as the $\hat{R}_y$ gate is sufficient to change the probability on each basis. It can create any 2-qubit state with real amplitudes as illustrated in Equation (9), representing any distribution with four data points $\vec{p}=[p_0,p_1,p_2,p_3]^\top$. The critic, parameterized with $\bm{\theta_c}$, is comparable to the discriminator in standard GAN\textsuperscript{\cite{42}}, except that the final sigmoid layer has been removed, resulting in an unbounded output. We construct the critic as a fully-connected NN implemented with Pytorch, which consists of an input layer, two hidden layers and an output layer with a node arrangement of 4-5-3-1. Leaky ReLU functions are used in both the input and hidden layers\textsuperscript{\cite{43}}. The real data is obtained by sampling from a classical random distribution. 
	
	During the training, the critic is fed both real and generated data, and the output is used to alternately train G and the critic. The optimization mission of HQC-GAN is
	
	\begin{equation}
		\mathop{min}\limits_{\bm{\theta_g}} \mathop{max}\limits_{\bm{\theta_d}} D_{\bm{\theta_d}}(G(\bm{\theta_g}))-D_{\bm{\theta_d}}(\bm{x})
		+\lambda \mathbb{E}_{p_{\bm{\hat{x}}}(\bm{\hat{x}})}[(||\nabla_{\bm{\hat{x}}}D_{\bm{\theta_{d}}}(\bm{\hat{x}})||_2-1)^2] 
	\end{equation}
	where $\bm{x}$ is the real data distribution, $p_{\bm{\hat{x}}}(\bm{\hat{x}})$ indicates the distribution of $\bm{\hat{x}}=\bm{x}+\zeta(G(\bm{\theta_g})-\bm{x})$ with $\zeta\sim U(0,1)$ and $\lambda$ is the gradient penalty coefficient. The introduction of the gradient penalty aims to meet the 1-Lipschitz continuity criterion required for WGAN\textsuperscript{\cite{40, 41}}. See Appendix B for more detailed experimental settings. 
	
	We used our chip to learn three distributions: the normal distribution with $X\sim \mathcal{N}(\mu=1.5,\sigma=1)$, the log-normal distribution with $X \sim \mathcal{LN}(\mu=0.5,\sigma=1)$ and the bimodel distribution being the superposition of $X\sim \mathcal{N}(\mu=0,\sigma=0.5)$ and $X\sim \mathcal{N}(\mu=2,\sigma=0.3)$. The real data, represented by the black solid lines in \textbf{Figure 4c-e}, are obtained by sampling 10,000 points and truncating them into the interval $[0,3]$. We trained the network in five rounds with varied initialization settings, each with 500 epoches. For each epoch, the critic is trained three times first, then G is trained once. 
	
	The blue histograms in \textbf{Figure 4c-e} exhibit the average value of learning results, which mimic the target distributions. \textbf{Figure 4f-h} show the evolution of the critic loss and the Kullback-Leibler divergence (KLD), with the solid line representing the mean value over 5 rounds and the shadow part indicating the STD. In traditional GANs, the KLD is commonly used to estimate the similarity of two distributions and is often employed as a convergence indicator. As noted, the critic loss and KLD have a comparable convergence tendency and finally approach 0. The final KLD value for all three distributions is less than 0.05, suggesting that the real distributions are successfully loaded.
	
	\subsection{Generate Compressed Images}
	As an important application, classical GANs have been utilized to generate realistic images that do not exist in the training dataset. In this application, nonlinearity is critical for capturing complicated patterns and nonlinear correlations in the data. Unfortunately, when a PQC is substituted for a classical G, the quantum G can only conduct linear transformations since quantum evolution is unitary. To introduce nonlinearity, one typical way is to trace the qubits out\textsuperscript{\cite{23,44,45}}, i.e., to add auxiliary qubits to the generator, and discard them during the measurement stage. However, this method is not suitable for our experiment since only two qubits are available. To address this challenge, we introduce a new method by placing a classical NN in front of the original quantum G, giving rise to a new hybrid G that performs well in the job of learning compressed MNIST images.
	
	\textbf{Figure 5a} depicts the training framework, which resembles the structure seen in \textbf{Figure 4a}. The significant distinction is the addition of a noise source $Z$ and the use of the hybrid G. The noise vector $\bm{z}$ is obtained by sampling from a uniform distribution within the range [0, 1]. As previously stated, the hybrid G is made up of two parts: one classical, which is a fully-connected NN, and the other quantum, which is a PQC. The incorporation of the classical NN enables straightforward nonlinear data transformations using activation functions. In this experiment, a $2\times2$ NN activated by Leaky ReLU functions is used. \textbf{Figure 5b} illustrates the quantum part implemented by our chip, which consists of an encoding layer and a training layer $U_{G}(\bm{\theta_q})$. The encoding layer occurs because the equation  $\hat{R}_y(\phi_1+\phi_2)=\hat{R}_y(\phi_1)+\hat{R}_y(\phi_2)$ holds true. To acquire generated data, the noise vector $\bm{z}$ is first input into the classic part of G, and undergoes a nonlinear transformation to become $G(\bm{\theta_c};\bm{z})$. The results are then encoded into the angle of $\hat{R}_y(z)$ gates and transformed into the output state $|G(\bm{\theta_c},\bm{\theta_q};\bm{z})\rangle=U_{G}(\bm{\theta_q})|G(\bm{\theta_c};\bm{z})\rangle$. Finally, measuring along the computational basis yields the generated data $G(\bm{\theta_c},\bm{\theta_q};\bm{z})$. 
	
	The training process is similar to the data loading task, except that in each epoch, a batch of data is sampled separately from the generated and real dataset, rather than a single distribution, to optimize G and D. The hyperparameters used in this experiment can be found in Appendix B. The optimization objective is
	
	\begin{equation}
		\mathop{min}\limits_{\bm{\theta_{g_c}},\bm{\theta_{g_q}}} \mathop{max}\limits_{\bm{\theta_{d}}} L(\bm{\theta_{g_c}},\bm{\theta_{g_q}},\bm{\theta_{d}})= \frac{1}{N}\sum_{i=0}^{N-1}[D_{\bm{\theta_{d}}}(G(\bm{\theta_{g_c}},\bm{\theta_{g_q}};\bm{z_i}))-D_{\bm{\theta_{d}}}(\bm{x_i})+\lambda\mathbb{E}_{p_{\bm{\hat{x}_i}}(\bm{\hat{x}_i})} (||\nabla_{\bm{\hat{x_i}}}D_{\bm{\theta_{d}}}(\bm{\hat{x}_i})||_2-1)^2]
	\end{equation}
	where $\bm{x_i}$ is a sample from the training dataset, and N is the batch size. The backpropagation approach is ineffective for optimizing $\bm{\theta_{g_c}}$ due to the quantum component between the classical NN and the output. Thus, we utilize the finite difference approach to get the approximate gradient, which is
	
	\begin{equation}
		d\bm{\theta_{g_c}}=\frac{L(\bm{\theta_{g_c}}+\epsilon)-L(\bm{\theta_{g_c}}-\epsilon)}{2\epsilon}
	\end{equation}
	where $\epsilon$ is a small incremental value.
	
	\begin{figure}[H]
		\includegraphics[width=1\textwidth]{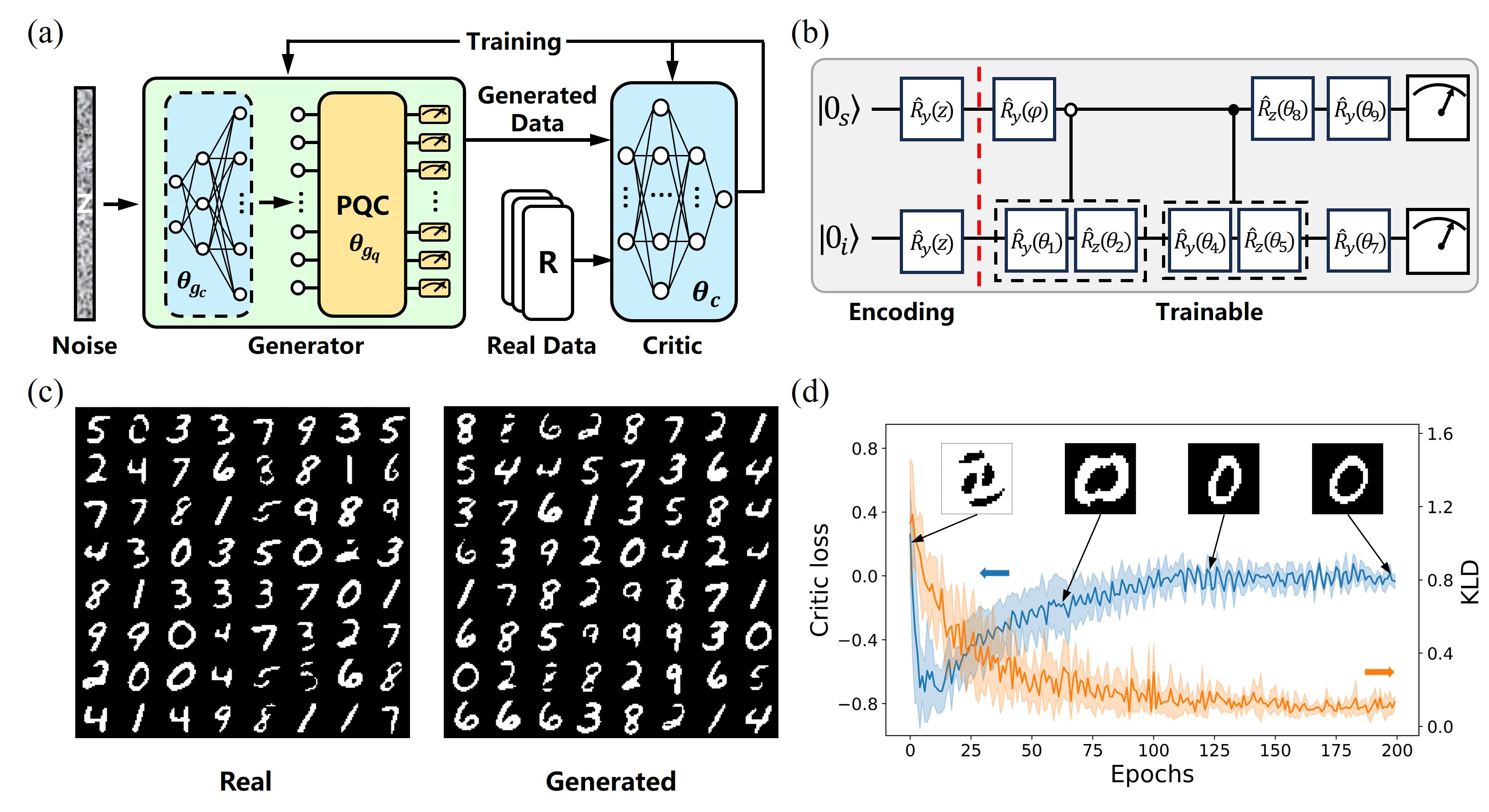}
		\caption*{\textbf{Figure 5.} a)The HQC-GAN framework for learning images, in which G is a combination of a NN and a PQC. The classical NN introduces nonlinearity for G by performing a nonlinear transformation on the input noise. b)The corresponding quantum circuit of $U(\bm{\theta_{g_q}})$, which includes an encoding layer and a training layer. c)Visualization of binarized MNIST images after IPCA, with real data examples on the left and chip-generated examples on the right. d)The progress of critic loss (blue) and KLD (orange) for digit 0, with the solid line and the shaded zone representing the average and STD of five runs with different initialization. The embedded images illustrate the evolution of the generated digit 0 as epochs proceed.}
		\label{fig:boat1}
	\end{figure}

	As a demonstration, we use this framework to train the chip to produce MNIST images. The MNIST dataset contains 60,000 handwritten images ranging from 0 to 9 with a size of $k=28\times28$, exceeding our chip's capabilities. Inspired by Ref. \cite{46,47}, we preprocess the pictures by compressing them to dimension $k=3$ with the principal component analysis (PCA) algorithm, resulting in the feature data $\vec{x}=[x_0,x_1,x_2]^\top$ for each compressed image. To match the image data with the quantum state represented in Equation (9), we augment $\vec{x}$ to ${x}'=[\vec{x}^\top,0.5]^\top$, and utilize the map $p_i=x_i'/\sum_{i=0}^{3}x_i'$, which assures a one-to-one mapping between $\vec{x}$ and $\vec{p}$, retaining all information.
	
	We then trained the chip to generate all nine digits (0–9) one at a time. The batch size is set to be $N=5$, and 5 rounds with each 200 epochs are conducted. To visualize the quality of the created images, we utilize the inverse PCA (IPCA) method to retrieve the data to the size of $28\times28$, and display the real and generated images with binarization in \textbf{Figure 5c}. The images are randomly selected from training and generated datasets. As can be seen, we produced images that are comparable in quality to the original. \textbf{Figure 5d} displays the curve of critic loss and KLD of digit 0, which converge to be approximately 0. The convergence and creation of high-quality images demonstrate the efficacy of our strategy. 
	
	\section{Discussion and Conclusion}
	
	Our scheme for generating arbitrary 2-qubit pure states can be extended to arbitrary 2-qudit by adopting high-dimensional encoding. In our strategy, AMZI is used to modulate the splitting ratio of photon pairs, resulting in an entangled state with controllable amplitude. In Ref. \cite{35}, the authors employ high-dimensional encoding to realize 4-dimensional $C\hat{U}$ gates, but the AMZI's phases are fixed to deterministic values for separating photon pairs, leaving only the maximum entangled state available. Our approach can be directly translated to the device by changing the AMZI step to obtain arbitrary 2-ququart pure states without any further changes. We predict that as the encoding dimension increases, our scheme's advantages will become more apparent. Furthermore, because the essence of our approach is to post-select photons in specific wavelengths through filters in order to change the quantum state amplitudes, it has a wider range of applications. In most quantum photonic computing chips, the first step is to generate entangled states\textsuperscript{\cite{17,48,49,50}}. Our approach can be considered an alternative solution for tuning the amplitude to perform more complex computations. In Ref. \cite{51}, the authors change the quantum state amplitude by tuning micro-ring filters, similar to our scheme, but applied in quantum networks.
	
	Our hybrid generator can alternatively be viewed as first preparing latent data with a conventional NN before inputting it into the quantum generator. This is distinct from existing hybrid generator structures, in which the quantum circuit prepares latent data and the classical neural network does post-processing, resulting in the creation of exclusively classical data\textsuperscript{\cite{24,25,26}}. In contrast, our hybrid generator maintains information in the quantum state, allowing for easier integration with subsequent quantum circuits. In this sense, it acts as a distribution loader to be a subset of larger algorithms\textsuperscript{\cite{52}}, but more advanced than the pure quantum version\textsuperscript{\cite{19,20}}. However, as mentioned before, the drawback is that due to the quantum layer between the NN and the measurement outcomes, backpropagation cannot be used, leading to time-consuming gradient acquisition for each parameter. As a result, the number of parameters in the classical NN must be carefully selected to minimize the runtime.
	
	It is acknowledged that due to the chip's limiting size, only four dimensions are available—which allows just conceptual demonstrations of quantum GANs. The few available qubits are also a challenge for other quantum computing platforms at present, but combining more advanced algorithms may help alleviate this issue. For instance, a variant of HQC-GAN called patch GAN has been proposed in Ref. \cite{23}, which combines the generated results from small-scale quantum subgenerators to represent large-scale data, opening up new prospects for the recent applications of quantum GANs. 
	
	In conclusion, we have designed a silicon quantum photonic chip made up of two spiral waveguides, two AMZIs, and 14 phase shifters that can generate any two-qubit pure state. The single chip is utilized for learning both quantum and classical data. Initially, we presented a PQ-GAN on this chip, training it to produce a high-fidelity single-qubit pure and mixed state. Then, we then integrated our quantum device with classical discriminators to show an HQC-GAN, which loaded three classical distributions onto the basis-state probabilities of two photonic qubits. Finally, we presented a hybrid generator that combines conventional NNs with quantum circuits for obtaining nonlinearity, and we used it to produce compressed MNIST images that are comparable in quality to the training set. Our work paves the way for the application of quantum photonic chips in generative learning. In the future, we expect to combine more advanced algorithms and larger-scale chips to implement quantum GANs, expecting to address more practical problems and applications in finance\textsuperscript{\cite{52}}, drug discovery\textsuperscript{\cite{25,26}} and more.

	\section*{Appendix A} 
	\textbf{Figure A1} shows a microscope photograph of the chip. The entire quantum chip has a dimension of 3mm $\times$ 0.8mm. The white lines and square pads act as electric conduits and metal pads for delivering electric signals to the PSs to manipulate the qubits. However, the PS $\theta_{11}$ was broken. We demonstrate that this has no impact on our experiments.
	
	As depicted in \textbf{Figure A2}, the transformations executed by the phase shifters $\theta_8$ through $\theta_{11}$ are $R_z(\theta_8)$, $R_y(\theta_9)$, $R_z(\theta_{10})$ and $U_{brok}$, respectively. Total transformation can be stated as
	
	\begin{equation}
		\tag{A1}
		U(\theta_8,\theta_9,\theta_{10})=U_{brok}R_z(\theta_{10})R_y(\theta_9)R_z(\theta_8)
	\end{equation}
	
	Since the combination of $R_z$ and $R_y$ gates can represent any 2-dimensional unitary, we can always find $\{\theta_{8}^0, \theta_9^0, \theta_{10}^0\}$ such that $R_z(\theta_{10}^0)R_y(\theta_9^0)R_z(\theta_8^0)=U_{brok}^{\dagger}$. Thus, we have
	
	\begin{equation}
		\tag{A2}
		I=U_{brok}R_z(\theta_{10}^0)R_y(\theta_9^0)R_z(\theta_8^0)=U_{brok}U_{brok}^{\dagger}
	\end{equation}	
	
	Assuming additional phase shifts $\{d\theta_8,d\theta_9\}$ is added, then
	
	\begin{equation}
		\tag{A3}
		\begin{aligned}
			&U(\theta_8^0+d\theta_8,\theta_9^0+d\theta_9,\theta_{10}^0)\\
			=&U_{brok}R_z(\theta_{10}^0)R_y(\theta_9^0+d\theta_9)R_z(\theta_8^0+d\theta_8)\\
			=&\underbrace{
				U_{brok}R_z(\theta_{10}^0)R_y(\theta_9^0)R_z(\theta_8^0)}_{I}
			R_z(-\theta_8^0)R_y(d\theta_9)R_z(\theta_8^0+d\theta_8)\\
			=&R_z(-\theta_8^0)R_y(d\theta_9)R_z(\theta_8^0+d\theta_8)
		\end{aligned}
	\end{equation}
	where $R_z(-\theta_8^0)$ has no physical meaning as it cannot be measured. By redefining $\theta_9 = d\theta_9$, $\theta_8 = \theta_8^0 + d\theta_8$ and ignoring $R_z(-\theta_8^0)$, the final transformation is obtained as
	
	\begin{equation}
		\tag{A4}
		U(\theta_8,\theta_9,\theta_{10}^0)=R_y(\theta_9)R_z(\theta_8)
	\end{equation}
	
	As a result, our experiment is unaffected by the broken phase shifter $\theta_11$. This is why, in our quantum circuit shown in \textbf{Figure 2b}, the final quantum gates for the signal photon qubit are $R_z$ and $R_y$.
	
	\begin{figure}[H]
		\includegraphics[width=1\textwidth]{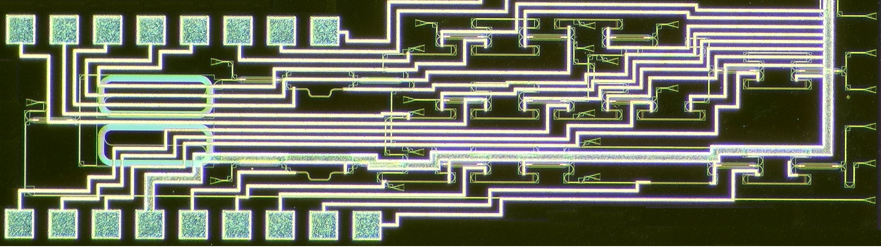}
		\caption*{\textbf{Figure A1.} The microscopic photograph of the chip.}
		\label{fig:boat1}
	\end{figure}
	
	\begin{figure}[H]
		\centering{\includegraphics[width=0.48\textwidth]{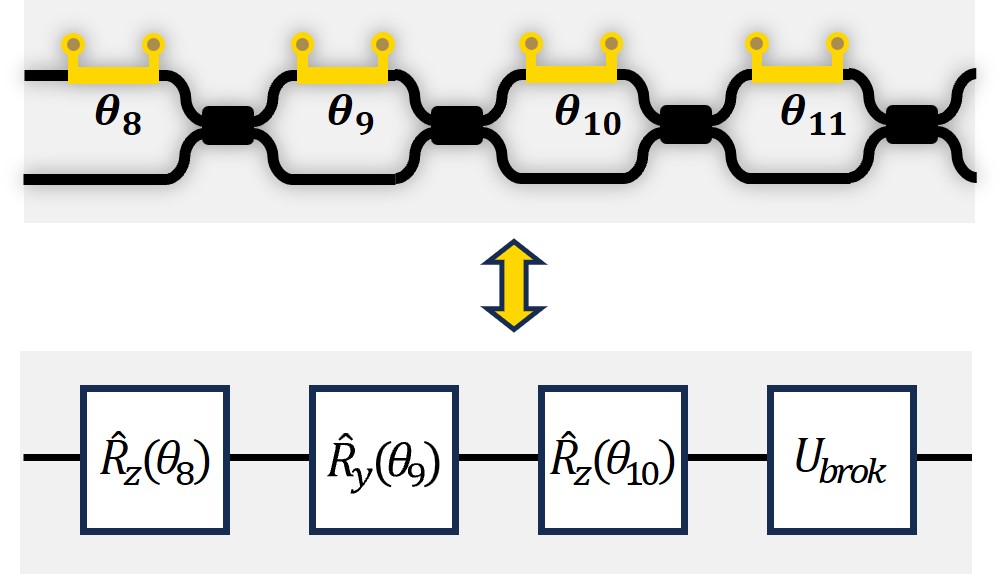}}
		\caption*{\textbf{Figure A2.} The connection between cascade phase-shifters and quantum gates, where $\theta_{11}$ is broken. }
		\label{fig:boat1}
	\end{figure}

	\section*{Appendix B} 
	In the experiments, our device implements the quantum component, while PyTorch is used to build the traditional NN. The three tasks are set up under various experimental circumstances. 
	
	In the single-qubit state learning problem, the parameters G and D are optimized using gradient descent (ascent) at learning rates of 0.02 and 0.1, respectively. For the distribution loading task, the learning rates for G and the critic are 0.08 and 0.1, respectively. In the compressed MNIST learning problem, the learning rates for G's classical portion, G's quantum part, and the critic are 0.02, 0.08, and 0.02, respectively.
	
	In the quantum state learning issue, a vanilla gradient is used and no optimizer is applied. The last two tasks employ the RMSProp optimizer\textsuperscript{\cite{53}} to update the parameters of both G and the critic, with optimizer hyperparameters set to $\beta=0.9$. The objective function has a gradient penalty coefficient of $\lambda=0.5$.
	
	The parameter-shift rule\textsuperscript{\cite{54}} is used to compute the gradients of the quantum portion in all three challenges, whereas PyTorch computes the gradients of the classical discriminator automatically. For the MNIST image learning issue, the gradients of G's classical portion are computed using finite differences, with an incremental value of $\epsilon=0.02$.

	\medskip
	\noindent\textbf{Acknowledgements} \par 
	\noindent This work was supported by the National Key Research and Development Program of China (Grants No. 2021YFB2800201), the National Natural Science Foundation of China (Grants No. U22A2082), the Ningbo Science and Technology Program (Grants No. 2023Z073), and the Provincial Natural Science Fundation of Zhejiang (Grants No. LD21F040003).
	
	\bibliographystyle{MSP}
	\bibliography{Reference}
	
\end{document}